# Brillouin optical time domain reflectometry for fast detection of dynamic strain incorporating double-edge technique

MINGJIA SHANGGUAN,[1,2,4] CHONG WANG,[2] HAIYUN XIA,[2,3,6] JIAWEI QIU,[2] GUOLIANG SHENTU,[1,4] XIANKANG DOU,[2,7] QIANG ZHANG,[1,4,8] AND JIAN-WEI PAN[1,4]

[1]*Shanghai Branch, National Laboratory for Physical Sciences at Microscale and Department of Modern Physics, USTC, Shanghai, 201315, China*
[2]*CAS Key Laboratory of Geospace Environment, USTC, Hefei, 230026, China*
[3]*Collaborative Innovation Center of Astronautical Science and Technology, HIT, Harbin 150001, China*
[4]*Synergetic Innovation Center of Quantum Information and Quantum Physics, USTC, Hefei 230026, China*
[5]*Jinan Institute of Quantum Technology, Jinan, Shandong 250101, China*
[6]*hsia@ustc.edu.cn*
[7]*dou@ustc.edu.cn*
[8]*qiangzh@ustc.edu.cn*

**Abstract**

For the first time, a direct detection BOTDR is demonstrated for distributed dynamic strain sensing incorporating double-edge technique, time-division multiplexing technique and upconversion technique. The double edges are realized by using the transmission curve and reflection curve of an all-fiber Fabry-Perot interferometer (FPI). Benefiting from the low loss of the fiber at 1.5 μm, the time-division multiplexing technique is performed to realize the double-edge technique by using only a single-channel FPI and only one piece of a detector. In order to detect the weak spontaneous Brillouin backscattering signal efficiently, a fiber-coupled upconversion detector is adopted to upconvert the backscattering signal at 1548.1 nm to 863 nm, which is detected by a Si-APD finally. In the experiment, dynamic strain disturbance up to 1.9 mε over 1.5 km of polarization maintaining fiber is detected at a sampling rate of 30 Hz. An accuracy of $\pm 30\,\mu\varepsilon$ and spatial resolution of 0.6 m is realized.

## 1. Introduction

In the past two decades, Brillouin-based distributed optical fiber sensors play an increasingly important role in various aspects of human life, such as large infrastructures, aerospace industry and geotechnical engineering [1]. The performance of distributed sensor based on Brillouin scattering has been improved significantly. Different techniques have been employed to design the distributed strain and temperature fiber sensors with long sensing range or with high spatial resolution. Recently, extensive research has been focused on developing distributed sensors for measurement of dynamic phenomena such as dynamic strain and sound wave.

The Brillouin-based distributed optical fiber dynamic strain sensor can mainly be divided into three types: the Brillouin optical time-domain analyzer (BOTDA), Brillouin optical correlation-domain analysis (BOCDA) and Brillouin optical time domain reflectometry (BOTDR).

The most widely used configuration is BOTDA. In BOTDA, a pulsed pump light and CW probe light with a frequency difference equivalent to the Brillouin frequency shift are launched at two opposite ends of the sensing fiber. The probe light at Stokes (anti-Stokes) frequency is amplified (attenuated) by the pump light through the Brillouin gain (loss) process. By scanning the frequency of either the pump light or probe light, the Brillouin spectrum is obtained [2]. The conventional BOTDA (C-BOTDA) has been reported for dynamic strain sensor over 120 m at a sampling rate of 3.9 kHz with a spatial resolution of 12 m and strain resolution of 256 με [3]. However, in order to obtain the Brillouin gain spectrum, the conventional BOTDA requires that the frequency difference between the probe and pump signal should be scanned to cover the whole Brillouin spectrum. To overcome this problem, slope-assisted BOTDA (SA-BOTDA) and Brillouin phase-shift optical time domain analysis (BPS-OTDA) have been proposed. By using SA-BOTDA, a dynamic strain sensor along 85 m fiber at a repetition rate of 400 Hz with a spatial resolution of 1.5 m was demonstrated [4]. A precision of ±20 με at 1.66 KHz sampling rate with 1 m resolution over a 160 m sensing fiber has been reported based on BPS-OTDA [5]. However, BOTDA is best reserved for monitoring single large disturbance, which induced Brillouin frequency shift is larger than the Brillouin bandwidth [6, 7]. Otherwise, the Brillouin spectrum at a special point, from which one retrieves the strain and temperature information, is influenced by potential transfer of power between the pulsed pump light and the counterpropagating CW probe light in the fiber from the point where the pulse enters and the point under estimation. Note that, this effect can be neglected for short-range measurement, but significant for a long range distributed measurement [8].

The principle of the BOCDA is based on modulating the frequency or phase of two counter-propagating CW lights to invoke stimulated Brillouin scattering at one section of the sensing fiber at any given time [9]. The BOCDA shows unique features of random accessibility of measuring position, high spatial resolution and high speed [10]. By using the BOCDA, a dynamic strain measurement along a 100 m fiber at a repetition rate of 20 Hz with an accuracy of ±50 με and a spatial resolution of about 80 cm has been demonstrated [11]. However, the sensing range in the classical BOCDA is about hundreds of the spatial resolution [10]. In order to enlarge the measurement range, additional time domain gating is demonstrated over 1 km sensing range with 7 cm spatial resolution [12]. But increasing the sensing range comes at the expense of achievable temporal resolution.

**Table 1. Performance chart of Brillouin-based dynamic strain sensors**

| Figure of merit / Detection scheme | Range | Sampling rate | Dynamic range | Spatial resolution | Strain precision | Reference |
|---|---|---|---|---|---|---|
| C-BOTDA | 120 m | 3.9 kHz | 256 με | 12 m | --- | [3] |
| SA-BOTDA | 85 m | 0.4 kHz | 600 με | 1.5 m | --- | [4] |
| BPS-OTDA | 160 m | 1.66 kHz | 2.56 mε | 1 m | ±20 με | [5] |
| BOCDA | 100 m | 20 Hz | 650 με | 0.8 m | ±50 με | [11] |
| C-BOTDR | 4 km | 2.5 Hz | --- | 10 m | --- | [14] |
| D-BOTDR(MZI) | 2 km | 1 Hz | 10 mε | 1.3 m | ±50 με | [15] |
| D-BOTDR(FPI) | 1.5 km | 30 Hz | 2.4 mε | 0.6 m | ±30 με | This work |

BOTDR is the simplest Brillouin-based technique, in which the variation of frequency and intensity of the spontaneous Brillouin backscatter signal is used to interrogate the temperature and strain along the sensing fiber [13]. Since the intensity of spontaneous Brillouin scattering is weak, it is challenging in developing a BOTDR with high temporal resolution. For example, by using a coherent-detection BOTDR (C-BOTDR), a maximum detectable frequency of 2.5 Hz over 4 km sensing fiber has been demonstrated. However, the proposed system has a poor spatial resolution of 10 m [14]. To overcome the slow scanning process of Brillouin spectrum, a direct-detection BOTDR (D-BOTDR) has been demonstrated by using an imbalanced Mach-Zhender interferometer (MZI), which is employed to convert the Brillouin frequency shift to intensity variation. A strain precision of $\pm 50 \mu\varepsilon$ over 2 km of sensing fiber with a spatial resolution of 1.3 m is demonstrated. However, the sampling rate is only 1 Hz [15]. The performances of typical Brillouin-based dynamic strain sensors are compared and listed in Table 1.

In conclusion, although extensive research has been focused on improving the Brillouin-based dynamic strain distributed fiber sensor, there are still challenges in developing a system with high sampling rate and high spatial resolution over long sensing fiber. In our previous works, the edge technique has been applied to measure the atmospheric wind speed in Doppler lidars [16, 17], and proposed to detect the Brillouin frequency shift induced by strain disturbance in fiber [18]. In addition, micro-pulse aerosol lidar [19] and all-fiber high spectral resolution wind lidar [20] have been constructed with upconversion technique. Especially, an upconversion BOTDR has been proposed and demonstrated to invoke the temperature profile over 9 km with spatial/temporal resolution of 2 m/15 s [21]. In this work, a direct-detection BOTDR based on double-edge technique is demonstrated by using a single-channel all-fiber Fabry-Perot interferometer (FPI) and a fiber-coupled upconversion single-photon detector, incorporating time-division multiplexing technique.

## 2. Principle

The double-edge technique is a powerful method to the detection of a small frequency shift. It usually uses two edge filters with opposite slopes [16, 17]. In this work, the transmission curve and reflection curve of the FPI is adopted to realize the double edges. Because of the symmetric arrangement of the double edges, the strain-induced Brillouin frequency shift will cause the signal to increase in one edge and to decrease in the other one. The variation of these two signals is sensitive to the change of the Brillouin spectrum, providing a unique detection scheme [18].

When a pulsed light launches into a sensing fiber, the power spectrum of the Brillouin backscatter $S_B(\upsilon)$ is a convolution of the pulse spectrum $S_P(\upsilon)$ and the Brillouin gain spectrum $g_B(\upsilon)$:

$$S_B(\upsilon) = S_P(\upsilon) \otimes g_B(\upsilon), \quad (1)$$

where $\otimes$ denotes the convolution. Due to the exponential decay of the acoustic waves in the fiber core, the Brillouin gain spectrum $g_B(\upsilon)$ has a Lorentzian spectral profile [22]:

$$g_B(\upsilon) = g_0 \left[1 + (\upsilon - \upsilon_B)^2 / (w_B/2)^2\right]^{-1}, \quad (2)$$

where $w_B$ is the full width at half-maximum (FWHM), $\upsilon_B$ is the Brillouin shift, $g_0$ is the peak value, given by

$$g_0 = 2\pi n_1^7 p_{12}^2 / c\lambda_p^2 \rho_0 V_a w_B, \quad (3)$$

where $n_1$ is the refractive index of the fiber, $p_{12}$ is the longitudinal elastic-optic coefficient, $c$ is the vacuum velocity of light, $\lambda_p$ is the pump wavelength, $\rho_0$ is the density, $V_a$ is the acoustic velocity within the fiber.

In Eq. (2), $\upsilon_B$ and $w_B$ can be expressed as a linear function of temperature $T$ and strain $\varepsilon$ as

$$v_B(T,\varepsilon) = v_B(T_0,0) + c_{v_B}^T(T-T_0) + c_{v_B}^\varepsilon \varepsilon, \qquad (4)$$

$$w_B(T,\varepsilon) = w_B(T_0,0) + c_{w_B}^T(T-T_0) + c_{w_B}^\varepsilon \varepsilon, \qquad (5)$$

where $T_0$ is the reference temperature, $c_{v_B}^T$, $c_{w_B}^T$ and $c_{v_B}^\varepsilon$, $c_{w_B}^\varepsilon$ are the temperature and strain coefficients for the Brillouin shift and spectrum broadening.

In this work, the pulse spectrum $S_p(v)$ can be approximated by a Gaussian function (this will be confirmed in the Experiment section), which can be expressed as:

$$S_p(v) = (\sqrt{\pi}\Delta v_M)^{-1} \exp\left[-(v-v_B)^2/\Delta v_M^2\right], \qquad (6)$$

where $\Delta v_M$ is the half-width at the 1/e intensity level of the pulse spectrum.

In order to eliminate the influence of the temperature on the measurement of the dynamic strain, the temperature profile along sensing fiber can be detected in advance. At a detected temperature $T_D$, the transmission of the Brillouin backscatter through the FPI varies with the strain of the sensing fiber experienced, which is given by

$$T_B(T_D,\varepsilon) = \int_{-\infty}^{\infty} S_B(v,T_D,\varepsilon)h(v)dv \Big/ \int_{-\infty}^{\infty} S_B(v,T_D,\varepsilon)dv, \qquad (7)$$

where $h(v)$ is the transmission function of the FPI. In this work, the cavity of the FPI is formed by two highly reflective multilayer mirrors that are deposited directly onto two carefully aligned fiber ends. The anti-reflection coated fiber inserted in the cavity provides confined light-guiding and eliminates secondary cavity. Since the FPI is fabricated with single-mode fiber with a negligible divergence in the cavity, its transmission is approximated to a Lorentzian function:

$$h(v) = T_M \Big/ \left[1+(v)^2/(\Delta v_{FPI}/2)^2\right], \qquad (8)$$

where $\Delta v_{FPI}$ is the FWHM of the transfer function, $T_M$ is the maximum transmission factor given by:

$$T_M = a_t(1-r_f)^2/(1-a_t \cdot r_f)^2, \qquad (9)$$

where $a_t$ is the attenuation factor, $r_f$ is the reflection coefficient of the reflecting ends.

For the purpose of improving the sensitivity of strain measurement and making full use of the Brillouin backscatter, the reflected signal of the Brillouin backscatter on the FPI is detected simultaneously, which can be expressed as:

$$R_B(T_D,\varepsilon) = \int_{-\infty}^{\infty} S_B(v,T_D,\varepsilon)r(v)dv \Big/ \int_{-\infty}^{\infty} S_B(v,T_D,\varepsilon)dv, \qquad (10)$$

where $r(v) = 1 - h(v)$ is the reflection function of the FPI.

Then, a response function is defined as

$$Q(T_D,\varepsilon) = \frac{a_0 T_B(T_D,\varepsilon) - R_B(T_D,\varepsilon)}{a_0 T_B(T_D,\varepsilon) + R_B(T_D,\varepsilon)}, \qquad (11)$$

where $a_0$ is a system constant, which can be determined in the calibration.

The operating principle of the double-edge is shown in Fig. 1(a). The double-edge technique is implemented by making use of the transmission curve and reflection curve of the FPI. As shown in Eq. (4) and Eq. (5), the strain not only affects the Brillouin frequency shift but also the bandwidth of the Brillouin gain spectrum, which lead to complement variations of the transmitted signal and reflected signal on the FPI. As shown in Fig. 1(b), with the known temperature and strain coefficients and temperature value (these coefficients can be found in Experiment section), the response function $Q$ varies monotonically with the strain that the sensing fiber experienced. In this work, a demonstrated experiment is carried out in the lab at temperature of 25 °C. The strain ranging from 0.7 mε to 1.9 mε is marked between two red points, as shown in Fig. 1(b).

### 3. Instrument

A schematic diagram of the direct-detection BOTDR is shown in Fig. 2. The laser adopts master oscillator power amplifier structure. The CW laser from a distributed feedback diode (DFB, 1548.1 nm) is chopped into a pulse train by using an electro-optic modulator (EOM$_1$, Photline, MXER-LN-10). The output pulse is amplified by an erbium-doped fiber amplifier (EDFA). The CW leakage is further suppressed to -50 dB with a same EOM$_2$, which minimize the amplified spontaneous emission (ASE) in the EDFA simultaneously. These two EOMs are driven and synchronized by using a pulse generator (PG), which determines the shape and pulse repetition frequency (PRF) of the laser pulse. The PRF of the pulses is set to 31.25 kHz, which indicates a maximum unambiguous detection range of 3.2 km. The ASE is further filtered by inserting a fiber Bragg grating (FBG) with ultra-narrow band of 6 pm. The filtered pulses are launched into the sensing fiber. The peak power of the pulse is 21 dBm. The sensing fiber comprised 1.5 km of unstrained polarization-maintaining fiber (PMF$_1$) with 1.8 m of strained fiber at the rear end. A brass ball with the weight of 110 grams is suspended at the end of the 1.8 m fiber, forming a pendulum. The 1.8 m strained region is isolated by using glue. When the pendulum oscillates about the equilibrium position, the strain of the sensing fiber experienced changes periodically.

The Stokes Brillouin backscatter signal (BBS) is picked out against the strong Rayleigh backscatter component by using a FBG$_2$ with bandwidth of 6 pm. The FPI is used as the frequency discriminator. Because the FPI is made of single-mode fiber, two polarization controllers (PC) are added at the front and rear ends of the FPI to eliminate the polarization-dependent loss. To guarantee the system stability, the FPI is cased in a temperature-controlled chamber (TCC). The transmitted signal of the Brillouin backscatter through the FPI is coupled into an upconversion detector, while the reflected signal of Brillouin backscatter is timely delayed through a 1.6 km PMF$_2$. As shown in Fig. 3, by using an optical switch (OS, Agiltron, NS-2x2), the transmitted and reflected signals can be coupled to the detector alternatively, incorporating a time division multiplexing (TDM) technique. The CW wave from the pump laser at 1950 nm is followed by a thulium-doped fiber amplifier (TDFA), which amplified the power of the pump laser up to 400 mW. The residual ASE noise is suppressed by using a 1.55/1.95 μm wavelength division multiplexer (WDM$_1$). The Brillouin backscatter signal and the pump laser are coupled into a periodically poled Lithium niobate waveguide (PPLN-W) via a WDM$_2$. It optimized quasi-phase matching condition is achieved by tuning the temperature of the PPLN-W with a thermoelectric cooler [23]. Here, the upconversion detector is integrated into an all-fiber module, in which the PPLN-W is coupled into a PMF/multimode fiber (MMF) at the front/rear end. The backscatter photons at 1548.1 nm are converted into sum-frequency photons at 863 nm and then picked out from the pump and spurious noise by using an interferometer filter (IF) at 863 nm with bandwidth of 1 nm. Finally, the photons at 863 nm are detected by a Si-APD. The TTL signals corresponding to the received photons are recorded on a digitizer (SPECTRUM, DN2.445-02, sampling rate of 500 MSa/s) and then transferred into a computer. Even though the conversion efficiency can approach a value larger than 99%, due to the coupling losses, the attenuation of the PPLN-W, and the limited quantum efficiency of the Si-APD at 863 nm, the final system efficiency of the UPD is 20% with a noise of 300 counts per second.

### 4. Experiment

Because the spatial resolution is determined by the width of a launched pulse, a short light pulse is required to improve the spatial resolution. As the pulse width becomes shorter, the bandwidth of the spectrum of Brillouin backscatter broadens, which leads to a larger dynamic range of the strain measurement. However, it will also result in the decreasing of the measurement sensitivity [24]. In this work, considering the trade-off between the spatial resolution and the measurement accuracy, a pulse light with the FWHM of 6 ns is used, which corresponds to the spatial resolution of 0.6 m. Figure. 4(a) presents the measured pulse in the time domain and transformed into the frequency domain, and Fig. 4(b) shows the output spectrum of the laser. A fast Fourier transform (FFT) is carried out with 502 points and sample-rate of 2.5 GSa/s. The result and its Gaussian fitting curve are shown as the inset in Fig. 4(a). The fitted bandwidth of the pulse shape in the frequency domain is 73.2 MHz, which agrees with $\Delta \upsilon \cdot \Delta t \approx 0.44$, indicating a transform limited optical pulse ($\Delta \upsilon / \Delta t$ is the temporal/ spectral width). As shown in Fig. 4(a), the assumption that the pulse spectrum is Gaussian function as shown in Eq. (6) is confirmed.

In a calibration experiment, a scheme of high spectral resolution BOTDR is used, which has been described in detail elsewhere [21]. In the FPI, a stacked piezoelectric transducer is used to axially strain the single-mode fiber inserted into the cavity. Frequency scanning of the FPI is achieved by scanning the cavity length.

As shown in Fig. 5(a), the transmission and reflection curves, from monochromatic CW laser, the laser pulse and BBS excited in the sensing fiber, are obtained by scanning the FPI over 600 MHz with a step of 15 MHz. By using the nonlinear fitting method, the response functions of CW laser (indicated by $Q_{CW}$), the laser pulse (indicated by $Q_{Pulse}$) and BBS (indicated by $Q_{BBS}$) are calculated as shown in Fig. 5(b) and compared in Fig. 5(d). The fitting residuals relative to the peaks are also provided for monitoring of the data quality, as shown in Fig. 5(c). Since the linewidth of the monochromatic CW laser (~3 KHz) is much less than that of the FPI, the measured spectrum represents the feature of the FPI. The fitting bandwidth of $Q_{CW}$ is 94 MHz. The fitted bandwidth of the response function of $Q_{BBS}$ is 180 MHz, which corresponds to the dynamic range of about 2.4 m$\varepsilon$.

By applying various strains on the sensing fiber at a constant temperature, the measured response functions are obtained. And then the nonlinear fitting method is performed to fit the calculated response functions to Voigt function. To gather statistics of the fitted Brillouin frequency shifts and spectrum broadenings at different strain, the strain coefficients are measured $c_{\upsilon_B}^{\varepsilon}$ =0.075 MHz/$\mu\varepsilon$ and $c_{w_B}^{\varepsilon}$ =0.05 MHz/$\mu\varepsilon$. Similarly, applying various temperatures on the unstrained sensing fiber, the temperature coefficients are measured $c_{w_B}^{T}$ =1.46 MHz/$\mu\varepsilon$ and $c_{w_B}^{T}$ =0.15 MHz/$\mu\varepsilon$. These parameters are used for calculating the response curve as shown in Fig. 1(b). Note that, the transmission and reflection curve of the BBS on the FPI in Fig. 5(a-3) is measured when the sensing fiber is free of strain and at the temperature of $25\,^{\circ}C$.

When the pendulum is displaced sideway from its resting (equilibrium position), it is subject to a restoring force due to gravity that will accelerate it back toward the equilibrium position. After released, the restoring force will make the brass ball oscillate about the equilibrium position, swinging back and forth. As the pendulum swings, the strain of the sensing fiber experienced changes periodically. And the period of the simple pendulum can be expressed as [25]:

$$P = 2(2L/g)^{1/2} \int_0^{\theta_0} 1/(\cos\theta - \cos\theta_0)^{1/2} d\theta, \qquad (12)$$

where $L$ is the length of the sensing fiber, g is the local acceleration of gravity, $\theta_0$ is the maximum angle that the pendulum swings away from vertical position (in experience, $\theta_0 = 55^{\circ}$).

As the pendulum swings, the tension of the sensing fiber shows the sinusoidal nature, as shown in Fig. 6. The sampling rate of the digitizer is set to 500 MSa/s, which corresponds to one sample every 0.2 m. In order to match the bandwidth (6 ns) of the laser pulse, a three-point average is performed, which leads to the spatial resolution of 0.6 m. The Brillouin backscatter signals from 1042 laser pulses are accumulated, which corresponds to the temporal resolution of 33.344 μs (namely 30 Hz sampling rate). After performing a low-pass FFT smoothing with cutoff frequency of 2 Hz to the retrieved strains, a nonlinear fitting method is carried out to fit the FFT smoothing results to a sine function, as shown in Fig. 6. The fitting period of the pendulum is 2.88 s, which is close to the calculated value (2.86 s, by using g=9.8m/s$^2$) from Eq. (12).

The scatterplot of the strain data from the FFT smoothing result and Sine fitting result is given in Fig. 7(a). The R-square and slope of the linear fit are 0.991 and 0.995 respectively. Figure. 7(b) is a histogram distribution of the stain difference between the results from the FFT smoothing and Sine fitting. The dash line superimposed on the histogram is a Gaussian fit with a mean of $1\,\mu\varepsilon$ and the FWHM of $60\,\mu\varepsilon$.

## 5. Conclusion

A direct detection BOTDR has been demonstrated for distributed dynamic strain sensing incorporating double-edge technique, time-division multiplexing technique and upconversion technique. The double edges are realized by using the transmission curve and reflection curve of an all-fiber FPI. Benefiting from the low loss of the fiber at $1.5\,\mu m$, the TDM is performed to realize the double-edge only using one single-channel FPI and a detector. By using upconversion technique, a high signal-to-noise upconversion detector is used to detect the weak spontaneous Brillouin backscatter signal efficiently. In demonstrated experiment, dynamic strain disturbance up to $1.9\,m\varepsilon$ over the 1.5 km of PMF is detected at a sampling rate of 30 Hz with an accuracy of $\pm 30\,\mu\varepsilon$ and spatial resolution of 0.6 m.

## Funding

This work was supported by National Natural Science Foundation (41274151,41421063); National Fundamental Research Program (2011CB921300, 2013CB336800);CAS Hundred Talents Program (D); CAS Program (KZZD-EW-01-1) and the 10000-Plan of Shandong Province; the Fundamental Research Funds for the Central Universities (WK6030000043).

## Acknowledgments

We thank Dr. Mingyang Zheng and Dr. Xiuping Xie for their help in manufacturing the upconversion detector.

## References

1. C. A. Galindex-Jamioy and J. M. López-Higuera, "Brillouin distributed fiber sensors: an overview and applications," J. Sens. 2012, 204121 (2012).
2. T. Kurashima, T. Horiguchi, and M. Tateda, "Distributed-temperature sensing using stimulated Brillouin scattering in optical silica fibers," Opt. Lett. 15(18), 1038–1040 (1990).
3. P. Chaube, B. G. Colpitts, D. Jagannathan, and A. W. Brown, "Distributed fiber-optic sensor for dynamic strain measurement," IEEE Sens. J. 8(7), 1067–1072 (2008).
4. Y. Peled, A. Motil, L. Yaron, and M. Tur, "Slope-assisted fast distributed sensing in optical fibers with arbitrary Brillouin profile," Opt. Express 19(21), 19845–19854 (2011).

5. J. Urricelqui, A. Zornoza, M. Sagues, and A. Loayssa, "Dynamic BOTDA measurements based on Brillouin phase-shift and RF demodulation," Opt. Express 20(24), 26942–26949 (2012).
6. T. R. Parker, M. Farhadiroushan, R. Feced, V. A. Handerek, "Simultaneous distributed measurement of strain and temperature from noise-initiated Brillouin scattering in optical fibers," Quantum Electron. 34, 645–659 (1998).
7. T. Horiguchi, K. Shimizu, T. Kurashima, M. Tateda, and Y. Koyamada, "Development of a distributed sensing technique using Brillouin scattering," J. Lightwave Technol. 13, 1296–1302 (1995).
8. L. Thévenaz, S. F. Mafang, and J. Lin, "Effect of pulse depletion in a Brillouin optical time-domain analysis system," Opt. Express 21(12), 14017–14035 (2013).
9. K. Hotate, "Fiber distributed Brillouin sensing with optical correlation domain techniques," Opt. Fiber Technol. 19(6), 700–719 (2013).
10. K. Y. Song, Z. He, and K. Hotate, "Distributed strain measurement with millimeter-order spatial resolution based on Brillouin optical correlation domain analysis," Opt. Lett. 31(17), 2526–2528 (2006).
11. K. Y. Song, M. Kishi, Z. He, and K. Hotate, "High-repetition-rate distributed Brillouin sensor based on optical correlation-domain analysis with differential frequency modulation," Opt. Lett. 36(11), 2062–2064 (2011).
12. K. Hotate, H. Arai, and K. Y. Song, "Range-enlargement of simplified Brillouin optical correlation domain analysis based on a temporal gating scheme," SICE J. Cont. Meas. Syst. Int. 1, 271–274 (2008).
13. H. H. Kee, G. P. Lees, and T. P. Newson, "All-fiber system for simultaneous interrogation of distributed strain and temperature sensing by spontaneous Brillouin scattering," Opt. Lett. 25(10), 695-697 (2000).
14. F. Wang, X. Zhang, X. Wang, and H. Chen, "Distributed fiber strain and vibration sensor based on Brillouin optical time-domain reflectometry and polarization optical time-domain reflectometry," Opt. Lett. 38(14), 2437-2439 (2013).
15. A. Masoudi, M. Belal, and T. P. Newson, "Distributed dynamic large strain optical fiber sensor based on the detection of spontaneous Brillouin scattering," Opt. Lett. 38(17), 3312–3315 (2013).
16. H. Xia, D. Sun, Y. Yang, F. Shen, J. Dong, and T. Kobayashi, "Fabry-Perot interferometer based Mie Doppler lidar for low tropospheric wind observation," Appl. Opt. 46(29), 7120–7131 (2007).
17. H. Xia, X. Dou, D. Sun, Z. Shu, X. Xue, Y. Han, D. Hu, Y. Han, and T. Cheng, "Mid-altitude wind measurements with mobile Rayleigh Doppler lidar incorporating system-level optical frequency control method," Opt. Express 20(14), 15286–15300 (2012).
18. H. Xia, C. Zhang, H. Mu, and D. Sun, "Edge technique for direct detection of strain and temperature based on optical time domain reflectometry," Appl. Opt. 48(2), 189-197 (2009).
19. H. Xia, G. Shentu, M. Shangguan, X. Xia, X. Jia, C. Wang, J. Zhang, J. S. Pelc, M. M. Fejer, Q. Zhang, X. Dou, and J. W. Pan, "Long-range micro-pulse aerosol lidar at 1.5 μm with an upconversion single-photon detector," Opt. Lett. 40(7), 1579–1582 (2015).
20. M. Shangguan, H. Xia, C. Wang, J. Qiu, G. Shentu, Q. Zhang, X. Dou, and J. W. Pan, "All-fiber upconversion high spectral resolution wind lidar using a Fabry-Perot interferometer," Opt. Express 24(17), 19322-19336 (2016).
21. H. Xia, M. Shangguan, G. Shentu, C. Wang, J. Qiu, M. Zheng, X. Xie, X. Dou, Q. Zhang, and J. W. Pan, "Brillouin optical time-domain reflectometry using up-conversion single-photon detector," Opt. Commun. 381, 37-42 (2016).
22. D. Heiman, D. S. Hamilton, and R. W. Hellwarth, "Brillouin scattering measurements on optical glasses," Phys. Rev. B. 19, 6583–6589 (1979).

23. G. Shentu, J. S. Pelc, X. Wang, Q. Sun, M. Zheng, M. M. Fejer, Q. Zhang, and J. Pan, "Ultralow noise up-conversion detector and spectrometer for the telecom band," Opt. Express 21(12), 13986–13991 (2013).
24. H. Naruse and M. Tateda, "Trade-off between the spatial and the frequency resolutions in measuring the power spectrum of the Brillouin backscattered light in an optical fiber," Appl. Opt. 38(31), 6516–6521 (1999).
25. G. Shentu, J. S. Pelc, X. Wang, Q. Sun, M. Zheng, M. M. Fejer, Q. Zhang, and J. Pan, "Ultralow noise up-conversion detector and spectrometer for the telecom band," Opt. Express 21(12), 13986–13991 (2013).
26. H. Naruse and M. Tateda, "Trade-off between the spatial and the frequency resolutions in measuring the power spectrum of the Brillouin backscattered light in an optical fiber," Appl. Opt. 38(31), 6516–6521 (1999).
27. F. M. S. Lima, and P. Arun. "An accurate formula for the period of a simple pendulum oscillating beyond the small angle regime," Am. J. Phys. 74(10), 892-895 (2006).

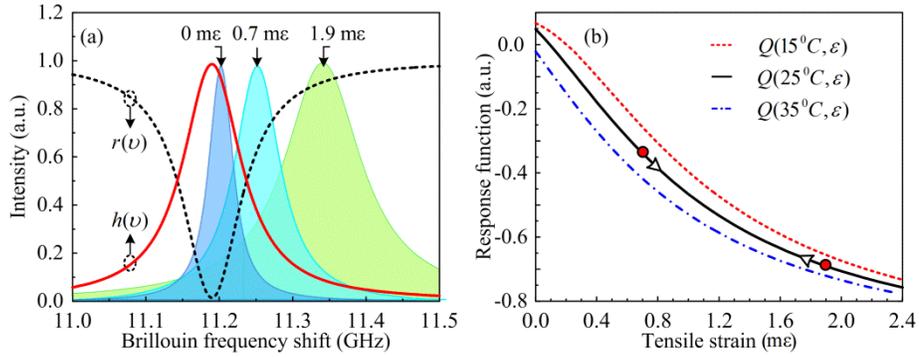

Fig. 1. (a): The transmission curve and reflection curve of the FPI, and the Brillouin gain spectrums when the sensing fiber experiences different strains at the temperature of 25 °C. (b) Response functions versus the tensile strain when the fiber experiences different temperatures.

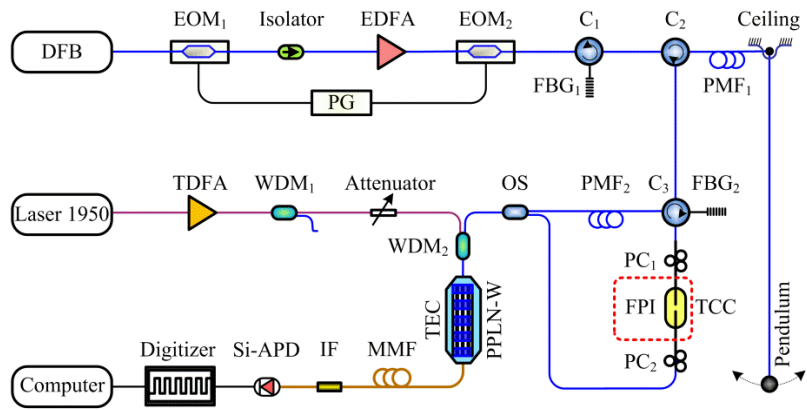

Fig. 2. Optical layout of the system. DFB, distributed feedback diode; EOM, electro-optic modulator; PG, pulse generator; EDFA, erbium doped fiber amplifier; C, circulator; FBG, fiber bragg grating; PMF, polarization-maintaining fiber; PC, polarization controller; FPI, Fabry-Perot interferometer; TCC, temperature controlled chamber; OS, optical switch; TDFA, thulium doped fiber amplifier; WDM, wavelength division multiplexer; PPLN-W, periodically poled Lithium niobate waveguide; TEC, thermoelectric cooler; MMF, multimode fiber; IF, interferometer filter.

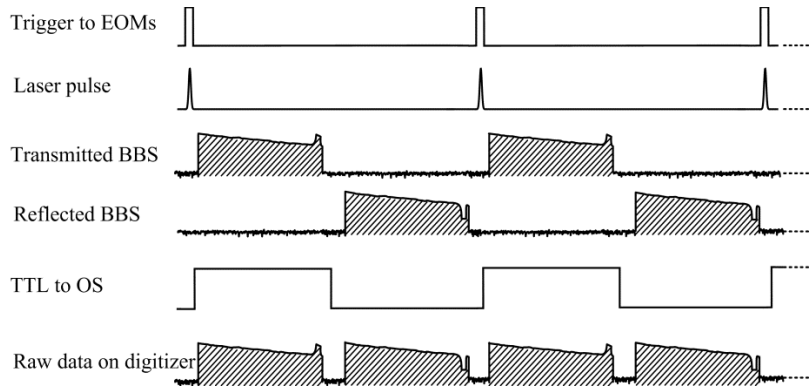

Fig. 3. Timing sequence of data acquisition

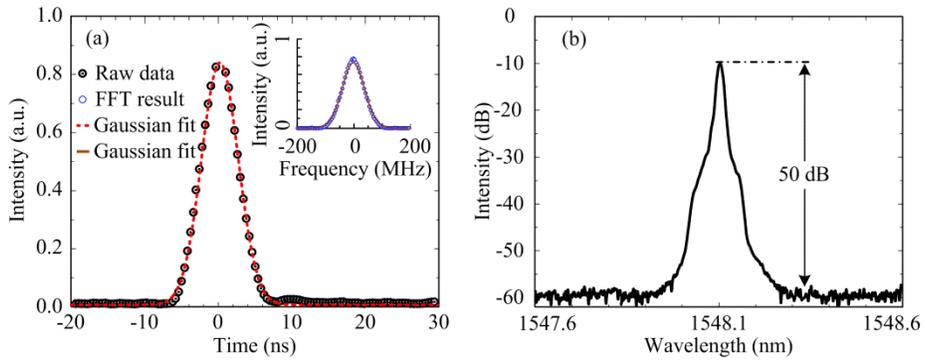

Fig. 4. (a): The pulse shape in time domain and in frequency domain (inset figure, obtain by Fourier transform of time profile of pulse), (b): the output spectrum of the laser.

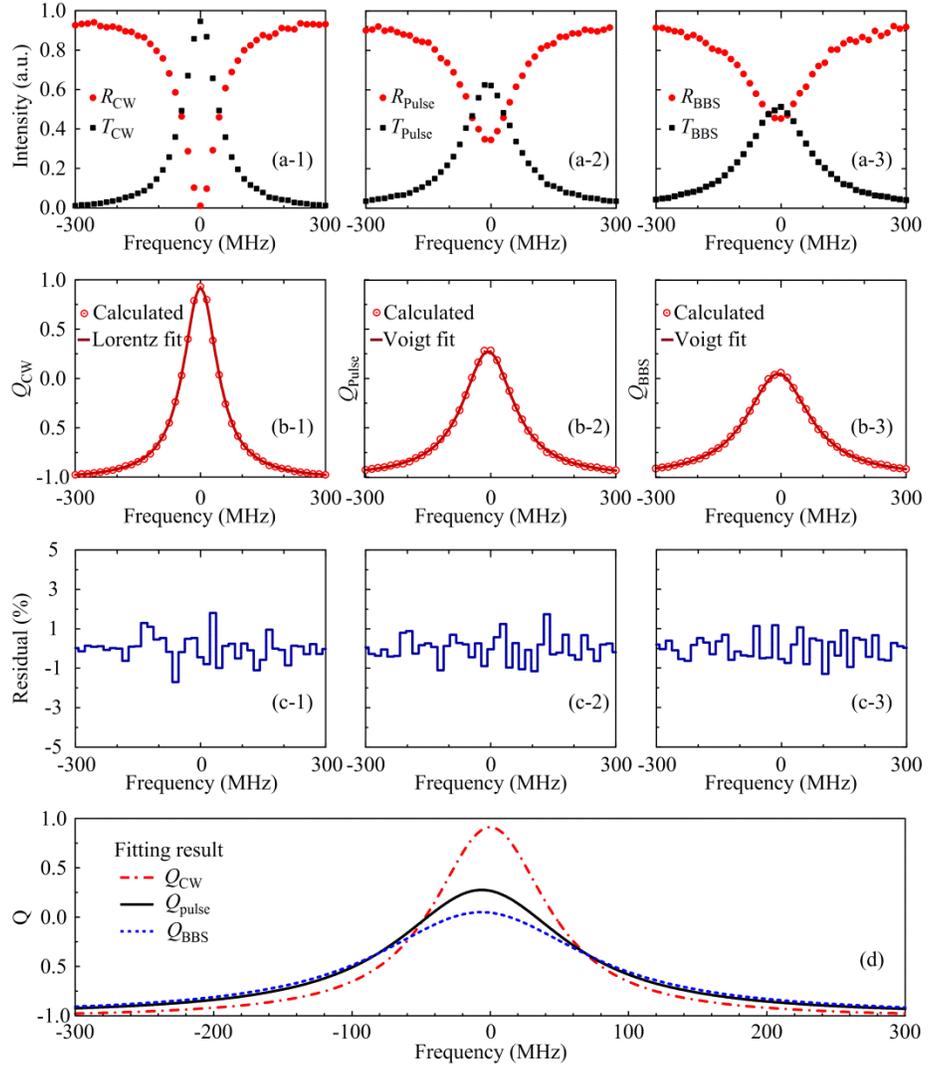

Fig. 5. (a): Transmission and reflection curves of a monochromatic CW laser (a-1, indicated by $T_{CW}$ and $R_{CW}$), the laser pulse with bandwidth of 6 ns (a-2, indicated by $T_{Pulse}$ and $R_{Pulse}$), and Brillouin backscatter of the laser with bandwidth of 6 ns (a-3, indicated by $T_{BBS}$ and $R_{BBS}$) by scanning the cavity length of the FPI, (b): the calculated response functions and its nonlinear fitting result, (c): residuals between the calculated response functions and its fitting results, which are relative to the peak, (d): comparison among three fitting response functions.

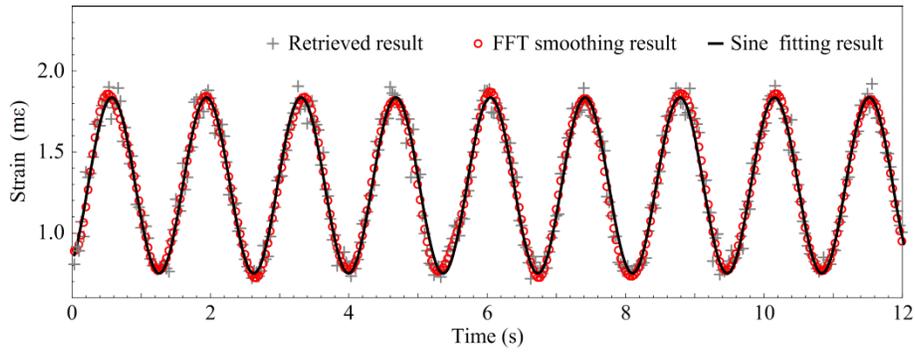

Fig. 6. Retrieved strain results with a repetition rate of 30 Hz, FFT smoothing result to the retrieved data, and Sine fitting result to the FFT smoothing result

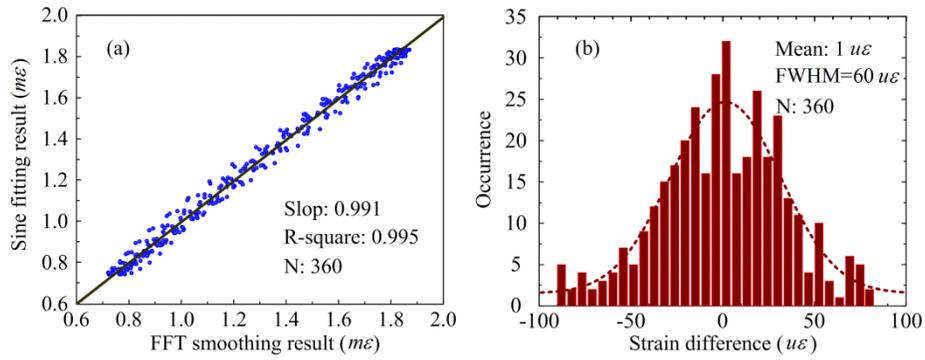

Fig. 7. (a): Scatterplot of strain data from FFT smoothing result and Sine fitting result, (b): Histogram distribution of the strain difference between the FFT smoothing result and Sine fitting result.